\documentstyle[12pt]{article}
\begin{document}

\newcommand{\PL}[3]{Phys. Lett. {\bf #1}, #2 (19#3)}
\newcommand{\PR}[3]{Phys. Rev. {\bf #1}, #2 (19#3)}
\newcommand{\PRL}[3]{Phys. Rev. Lett. {\bf #1}, #2 (19#3)}
\newcommand{\douba}[2]{\mbox{$
\left( \begin{array}{c} #1    \\ #2  \end{array} \right)$}}
\newcommand{\doub}[3]{\mbox{$
\left( \begin{array}{c} #1    \\ #2  \end{array} \right)_#3$}}
\newcommand{\stacksub}[2]{\ _{\stackrel{\textstyle
#1}{\scriptstyle #2}}\ }
\def\mxth{\mathsurround=0pt }
\def\xversim#1#2{\lower2.pt\vbox{\baselineskip0pt \lineskip-.5pt
  \ialign{$\mxth#1\hfil##\hfil$\crcr#2\crcr\sim\crcr}}}
\def\simgr{\mathrel{\mathpalette\xversim >}}
\def\simle{\mathrel{\mathpalette\xversim <}}
\def\Tr{{\rm Tr}}
\def\etc{{\it etc}}
\def\ra{\rightarrow}
\def\x{\times}
\def\beq{\begin{equation}}
\def\eeq{\end{equation}}
\def\beqa{\begin{eqnarray}}
\def\eeqa{\end{eqnarray}}
\def\L{{\cal L}}
\def\ie{{\it i.e.}}

\titlepage

\vspace{8ex}

\begin{center} \bf

PROTON DECAY \\   \rm

\vspace{4ex}

PAUL LANGACKER \\ University of Pennsylvania\\ Department of Physics
\\ Philadelphia, Pennsylvania  19104-6396 \\

\vspace{6ex}

ABSTRACT
\end{center}

The status of proton decay is described, including general motivations
for baryon number violation, and the present and future experimental
situation.  Grand unification with and without supersymmetry is
considered, including possible evidence from coupling constant
unification and implications for proton decay and neutrino mass.

\section{Motivations}

Baryon number is almost certainly not absolutely conserved.

\begin{itemize}
\item There is no compelling
reason to think that baryon number {\it is} conserved.
The only convincing mechanism we have for
ensuring an absolute conservation law is gauge invariance.  For
example, electromagnetic gauge invariance guarantees that electric charge
is conserved and implies the existence of a massless photon.  However,
there is no analogous baryton -- i.e., there is no long range force
coupling to $B$.  We know from E\"{o}tvos-type experiments~\cite{p1}
that if there were such a gauge boson its coupling would have to be
incredibly small, $g^2_B/4\pi < 6 \x 10^{-48}$.  Hence, baryonic gauge
invariance cannot be invoked and there is no good reason to suspect
absolute conservation.

\item Black holes do not remember baryon
number.  If a proton were to drop into a black hole its
quantum numbers would disappear from the universe, violating baryon
number.

\item It has been known for some time~\cite{p2} that baryon number is
violated in the weak interactions via the weak anomaly, as shown in
Figure~\ref{fig1}.  The idea is that the vacuum state is not unique
and there are degenerate vacua characterized by different values for
$B$.  There are nonperturbative tunnelling amplitudes to make
transitions from one vacuum to another.  However, these are incredibly
slow, characterized by rates proportional to ${\rm exp} (-4 \pi \sin^2
\theta_W/\alpha) \sim 10^{-172}$, and are irrelevant for proton decay.
However, it has been realized and emphasized recently~\cite{p3} that
thermal fluctuations at the time of the electroweak phase transition
could lead to transitions, and this has significant implications for
the baryon asymmetry of the universe.
\begin{figure}
\vspace{5cm}
\caption{Schematic diagram of baryon number violation in the weak
interactions.  There are degenerate vacua of different baryon number
and a possibility of tunnelling between them.}
\label{fig1}
\end{figure}

\item The baryon asymmetry of the universe~\cite{p4} is the small
difference between the number of baryons and antibaryons, $(n_B -
n_{\bar{B}})/ n_\gamma \sim 10^{-10}$.  This most likely implies $B$
violation at some time in the history of the universe.  It is also
possible that there was a small initial asymmetry or that there is a
large-scale separation between baryons and antibaryons, but these seem
to me to be much less likely.

\item Finally, it is easy to construct grand unification~\cite{p5,p6}
or other interactions which involve $\Delta B \neq 0$.
\end{itemize}

All of these should be viewed as reasonable motivations to consider
the possibility that $B$ is not conserved and that the proton is not
stable.

\section{Proton Decay Experiments}

It was possible by ``simple considerations'' to determine the limit
$\tau_p > 10^{20-22} \; yr$ in the early 1950's \cite{reines}.
Dedicated experiments, either looking for the disappearance of a
nucleon from the nucleus or searching directly for the decay products,
improved the limits to $\sim 10^{29}\; yr$ by the mid 1970's.  A
detailed review can be found in \cite{science}.

The modern experiments have pushed the limits still further.  Recent
experiments that have been completed, are still running, or are under
construction are listed in Table~\ref{tab1}.  A recent review is given
in \cite{barloutaud}.
\begin{table}   \centering
\begin{tabular}{|lllc|} \hline
exp & Type & Status & \parbox[t]{12em}{Sensitivity \ (kT yr)} \\
\hline
Kolar & Track. Cal. & running & $0.8$ \\
IMB & Water Cer. & stopped & $3.8 - 7.2$ \\
NUSEX & Track. Cal. & running & $0.36$ \\
HPW & Water Cer. & stopped (85) & $0.14$ \\
Kamiokande & Water Cer. & running & $3.76$ \\
FREJUS & Track. Cal. & stopped (88) & $1.5 - 2$ \\
Soudan 2 & Track. Cal. & - running (0.7 kT) & $0.94 (\ra 6)$ \\
\ & \ & (1\  kT completed '93) & \ \\ \hline
Super-Kamiokande & Water Cer. & approved & \ \\
 \ & \ & (22 kT completed '96) & $(\ra 100)$ \\ \hline
\end{tabular}
\caption{Modern proton decay experiments.  The last column shows their
sensitivity in kiloton-years. From~\protect\cite{barloutaud}.}
\label{tab1}
\end{table}

No experiment has observed proton decay.  There are some candidate
events in some of the experiments, but these could be due to neutrinos
interacting in the detector.  The present limits on two particularly
interesting modes are
\beqa e^+\pi^0 : && \tau/B > 10^{33} \; yr  \nonumber \\
\bar{\nu} K^+: && \tau/B > 10^{32} \; yr. \eeqa
These are interesting in ordinary and supersymmetric grand unified
theories, respectively.  In the future the $e^+ \pi^0$ limit will be
improved to $\sim 10^{34}\;yr$.

\section{Grand Unification}

Many of the problems of the standard model can be solved in a grand
unified theory~\cite{p5,p6}.  The basic idea is that the strong, weak,
and electromagnetic interactions are unified, \ie, embedded in a
simple group $G$.  If one could probe the theory at a large momentum
scale $Q \geq M_X$ for which symmetry breaking can be ignored one
would observe a single coupling constant.  In practice $G$ is broken
to the standard model group,\footnote{Or to a subgroup of $G$ which
contains the standard model.} $G \stacksub{\rightarrow}{M_X} SU_3 \x
SU_2 \x U_1$, at some large scale $M_X$ so that at lower energies the
three interactions appear different.  One of the predictions is that
the coupling constants that we observe at low energy, when run
theoretically up to high scales, should meet at $M_X$.  Grand
unification is a elegant idea but it does not incorporate gravity.

In addition to the coupling constants the $q$, $\bar{q}$, $l$,
$\bar{l}$ are all unified.  That is, they are related in the same
multiplets.  This typically explains charge quantization, \ie, why the
the atom is electrically neutral, and implies new interactions which
mediate proton decay.

The simplest grand unified theory is the Georgi-Glashow $SU_5$
model~\cite{p5}.  The fermions are placed in a complicated reducible
representation, consisting of a 5-plet,
\beq 5^*:\ \ \left( \begin{array}{c} \nu_e \\ e^- \end{array} \right)_L
\ \ \bar{d}_L ,\eeq
and a 10-plet,
\beq 10: \ \  e^+_L \ \ \left( \begin{array}{c} u \\ d \end{array} \right)_L
\ \ \bar{u}_L ,\eeq
for each fermion family.  One expects that the group will break
\beq SU_5 \stacksub{\ra}{M_X} SU_3 \x SU_2 \x U_1
\stacksub{\ra}{M_Z} SU_3 \x U_{1Q} \eeq
to the standard model.  There are new colored gauge bosons $X$ and $Y$
with charges $4/3$ and $1/3$ which mediate transitions between quarks
and leptons or between quarks and antiquarks.  These can mediate
proton decay, as shown in Figure~\ref{fig2}.  One expects the lifetime
to be of order
\beq \tau_{p \ra e^+ \pi^0} \sim \frac{M^4_X}{\alpha^2_5 m_p^5} \simgr
10^{30} \; yr \;{\rm [1980]} ,\eeq
where the experimental limit is from 1980.  Thus, one required
\beq M_X \simgr 10^{14} \; GeV \; {\rm [1980]} \eeq
to avoid too fast proton decay, with a grand desert between $M_X$ and
$M_Z$.  This is an enormous mass scale compared to those of the
standard model -- it was a daring extrapolation.  Still, $M_X$ is
sufficiently small compared to the Planck scale, $M_P = G_N^{-1/2}
 \sim 10^{19}\;
GeV$, that the partial unification without gravity is consistent.
\begin{figure}
\vspace{5cm}
\caption{Diagram for proton decay in the Georgi-Glashow $(SU_5)$ and
similar models.}
\label{fig2}
\end{figure}

One can also consider larger grand unified theories, in which more
particles are related by the symmetry.  For example, there is the
$SO_{10}$ model, $SU_5 \subset SO_{10}$, in which each family is
placed in a 16-plet, $16 = 5^* + 10 + 1$, which includes a new
neutrino $\bar{N}_L$, which may be superheavy.  There is also the even
larger group $SO_{10} \subset E_6$, in which each family is placed in
a $27 = 16 + 10 + 1$, which consists of the 16-plet, a heavy down-type
quark $(D_L \; , \; \bar{D}_L)$ which has no charged current weak
interactions (it is an $SU_2$ singlet),
a heavy lepton doublet  \doub{E^+}{E^0}{L}
\doub{\bar{E}^0}{E^-}{L} in which both the left and right-handed
components transform as weak doublets, and an additional (possibly
superheavy) neutrino $S_L$.

Grand unified theories have many interesting prediction.  In addition
to the coupling constant unification, charge quantization, and proton
decays, some of the simplest give correct predictions for the ratio
$m_b/m_\tau$.  Also the baryon number violating interactions could
generate a baryon asymmetry of the universe.  However, as alluded to
earlier, such an asymmetry could be later erased at the time of the
electroweak transition unless the GUT asymmetry had a non-zero
$B-L$~\cite{r1}.

\section{Supersymmetry}

Supersymmetry (SUSY) is a new type of symmetry which relates fermions
and bosons~\cite{r2}.  No known particles can be partners, so
supersymmetry requires a doubling of the particle spectrum.  There are
two major motivations for considering such schemes.  One is the
problem of the Higgs mass renormalization.  In the standard model the
loop diagrams in Figure~\ref{fig3} lead to
\beq m_H^2 = m^2_{H^0} + \delta m^2 ,\eeq
where  $m_{H^0}$ is the bare mass and
\beq \delta m^2 = O (g^2, \; \lambda,\; h^2)\Lambda^2 \eeq
is the correction.  The diagrams are quadratically divergent, so that
$\Lambda$ is a cut off.  In practice it should be viewed as the scale
at which new physics comes in to turn off the standard model and
regulate the integrals.  For example, if the next scale in nature is
the gravity (Planck) scale $M_P = 10^{19} \; GeV$ one requires
an enormous fine-tuning to ensure that the physical mass is of order 1
TeV or less.

\begin{figure}
\vspace{5cm}
\caption{One-loop diagrams contributing to the renormalization of the
Higgs mass.}
\label{fig3}
\end{figure}

There are two known ways to avoid this fine-tuning.  One is to
replace the elementary Higgs fields by some sort of dynamical symmetry
breaking.  However, this approach leads to other difficulties and no
satisfactory models have been constructed.  The other is to introduce
supersymmetry.  In this case the new particles of the theory enter the
corrections with opposite signs, so that
\beq m_H^2 = m^2_{H^0} + \delta m^2 - \delta m^2_{\rm SUSY} \eeq
If supersymmetry were exact there would be a complete cancellation.
Since supersymmetry has not yet been observed it must be broken, with
the new superpartners heavy.  If there is a soft breaking of the
supersymmetry the quadratic divergences still cancel, but one expects
to have finite remaining terms $O(g^2, \; \lambda, \; h^2) (m^2 -
\tilde{m}^2)$, where $m$ and $\tilde{m}$ represent the masses of an
ordinary particle and its superpartner.  We must therefore choose
\beq |\tilde{m}| \simle O(TeV) \eeq
as the typical mass scale of the new supersymmetric particles
to ensure that the Higgs mass is not too much larger than the desired
electroweak scale.  Supersymmetry predicts a rich structure of new
particles.  In addition to the superpartners there must be two Higgs
doublets (and their partners) instead of one.  All of the new
particles may be in the hundreds of GeV range where they will be
difficult to observe.  However, the motivation for supersymmetry
requires that they cannot be too heavy.  They can be searched for at
high energy colliders, and there may be indirect indications from low
energy precision experiments.

Another motivation for supersymmetry is gravity.  When one has a
gauged SUSY there is necessarily a spin-3/2 gauge particle known as
the gravitino, which is the superpartner of a spin-2 graviton.
Supergravity theories automatically bring gravity into the game.
However, they are not by themselves renormalizable.

One can consider supersymmetry either with or without grand
unification.  Superstring theories assume that the low energy theory
is supersymmetric.

\section{Unification of Coupling Constants}

Now let us consider the running of the scale-dependent effective
coupling constants, using the two-loop renormalization group equations
\beq \frac{d\alpha^{-1}_i}{d \ln \mu} = - \frac{b_i}{2 \pi} -
\sum^3_{j=1} \frac{b_{ij} \alpha_j}{8\pi^2} ,\eeq
where $\alpha_i = g_i^2/4\pi$, $i = 1,  2,  3$, is the coupling of
the $SU_3$, $SU_2$, or $U_1$ groups, respectively.  These equations
can be solved to yield
\beq  \alpha^{-1}_i (\mu) = \alpha_i^{-1} (M_X) - \frac{b_i}{2\pi}
\ln \left( \frac{\mu}{M_X} \right)
   + \sum^3_{j=1} \frac{b_{ij}}{4\pi b_j} \ln \left[
\frac{\alpha^{-1}_j (\mu)}{\alpha_j^{-1} (M_X)} \right] ,\eeq
where $\mu$ is an arbitrary momentum and $M_X$ is an arbitrary
reference point.  To an excellent first approximation one can neglect
the last (2-loop) term, in which case the inverse coupling constant
varies linearly with $\ln \mu$.  The 2-loop terms are small but not
entirely negligible.

In a grand unified theory one expects that the three couplings will
meet at $M_X$~\cite{r3}--\cite{r8a}, up to threshold
corrections~\cite{r9}, as shown in Figure~\ref{fig4}:
\begin{figure}
\vspace{5cm}
\caption{Running of the inverse coupling constants.  They can be
measured at low energies and extrapolated theoretically in a grand
unified theory.  They should meet at the unification scale up to small
threshold corrections.}
\label{fig4}
\end{figure}
\beq \alpha^{-1}_i (M_X) = \alpha_G^{-1} (M_X) + \delta_i + \Delta_i.
\eeq
The $\delta_i$ are small corrections associated with the low energy
threshold, from effects such as $m_t > M_Z$ and the non-degeneracy of
the new particles, $m_i^{\rm new} \neq M_{\rm SUSY}$.  Similarly,
there may be corrections $\Delta_i$ associated with $m_{\rm heavy}
\neq M_X$ at the high scale, or with non-renormalizable operators.

If one knows the values of the couplings at low energy then they can
be extrapolated theoretically   in terms of calculable coefficients.
The 1-loop terms are
\beq b_i = \left(\begin{array}{c} 0 \\ -\frac{22}{3} \\ - 11
\end{array} \right) + F \left(\begin{array}{c} \frac{4}{3} \\
\frac{4}{3} \\ \frac{4}{3} \end{array} \right) + N_H
\left(\begin{array}{c} \frac{1}{10} \\ \frac{1}{6} \\ 0 \end{array}
\right) ,\eeq
assuming the standard model. $F$ is the number of fermion families and
$N_H$ is the number of Higgs doublets.  In the minimal supersymmetric
extension of the standard model (MSSM), in which one has the minimal
number of new particles,
\beq b_i = \left(\begin{array}{c} 0 \\ -6 \\ - 9 \end{array} \right) +
F \left(\begin{array}{c} 2 \\ 2 \\ 2 \end{array} \right) + N_H
\left(\begin{array}{c} \frac{3}{10} \\ \frac{1}{2} \\ 0 \end{array}
\right) ,\eeq
where the difference is due to the fact that additional particles
enter the loops.  The 2-loop coefficients can be found in
\cite{r4}.

\section{Couplings at $\bf M_Z$}

To test the unification one must know the values of the couplings at
low energy.  We define the couplings $g_s = g_3$, $g=g_2$, and $g' =
\sqrt{3/5} g_1$ of the standard model $SU_3 \x SU_2 \x U_1$ group, and
the fine-structure constants $\alpha_i = g^2_i/4\pi$.  The extra
factor in the definition of $g_1$ is a normalization
condition~\cite{r3}.  The couplings are expected to meet only if the
corresponding group generators are normalized in the same way.
However, the standard model generators are conventionally normalized
as $\Tr (Q^2_s) = \Tr (Q^2_2) = 5/3 \Tr (Y/2)^2$, so the factor
$\sqrt{3/5}$ is needed to compensate.  The couplings are related to
$e$, the electric charge of the positron, by $e = g \sin \theta_W$,
where the weak angle is
\beq \sin^2 \theta_W = \frac{g^{'2}}{g^2 + g^{'2}} = \frac{g^2_1}{
\frac{5}{3} g^2_2 + g_1^2} \stacksub{\ra}{g_1 = g_2}
\frac{3}{8}.\eeq
One expects $\sin^2\theta_W = 3/8$ at the unification scale~\cite{r3}
for which $g_1 = g_2$.

Precision experiments determine the weak angle very well.  For
comparing with grand unification it is convenient to use the modified
minimal subtraction $(\overline{MS})$ renormalization scheme.
Experimentally\footnote{This value is obtained assuming the standard
model. In the MSSM, one has
$\sin^2\hat{\theta}_W (M_Z) = 0.2324 \pm 0.0006$. The difference is due
to the lower value (50-150 GeV) expected for the mass of the lightest
Higgs scalar in the MSSM. (The second Higgs doublet and the superpartners
do not significantly affect the precision observables.)}
\beq \sin^2\hat{\theta}_W (M_Z) = \frac{\alpha_1 (M_Z)}{\frac{5}{3}
\alpha_2 (M_Z) + \alpha_1 (M_Z)} = 0.2325 \pm 0.0007,\eeq
where most of the uncertainty is from the top quark mass.
Similarly,
\beq \alpha^{-1} (M_Z) = 127.9 \pm 0.2 = \frac{\alpha^{-1}_2
(M_Z)}{\sin^2 \hat{\theta}_W (M_Z)}\eeq
is obtained by extrapolating the observed electromagnetic fine
structure constant from low energies, where it is measured precisely,
to $M_Z$.  Much of the uncertainty is associated with the low energy
hadronic contribution to the photon vacuum
polarization\footnote{These same hadronic uncertainties lead to the major
theoretical uncertainty in $g_\mu -2$ and in the relationship between
$\sin^2 \hat{\theta}_W - M_Z$.} and some is from $m_t$.

For the strong coupling I will use
\beq \alpha_s (M_Z) = 0.12 \pm
0.01.  \label{alsv} \eeq
This value is somewhat higher than has been used in the
past.  It is motivated by a new theoretical analysis of event shapes
at LEP which yields $0.123 \pm 0.005$, as shown in Table~\ref{tab2}.
\begin{table} \centering

 \begin{tabular}{|c c|} \hline
 source & $\alpha_{s}(M_{Z})$   \\ \hline \hline
 $R_{\tau}$
 &$0.118 \pm 0.005$
 \\ \hline
 $DIS$
 & $0.112 \pm 0.005$
 \\ \hline
 $\Upsilon$, $J/\Psi$
 & $0.113 \pm 0.006$
 \\ \hline
 LEP($R$)
 & $0.133 \pm 0.012$
 \\ \hline
 LEP(events)
 & $0.123 \pm 0.005$
 \\ \hline
 average
 & $0.120 \pm 0.010$
 \\ \hline
 \end{tabular}

\caption[]{Values of $\alpha_{s}(M_{Z})$,
  adapted from \cite{r10}. $R_{\tau}$
  refers to the ratio of hadronic to leptonic $\tau$ decays; $DIS$ to
  deep-inelastic scattering; $\Upsilon$, $J/\Psi$ to onium decays;
  and LEP($R$) to the ratio of hadronic to leptonic $Z$ decays.
  LEP(events) refers to the event topology in $Z \longrightarrow$ jets.}
\label{tab2}
\end{table}
Previous analysis of event shapes using the same data gave the
somewhat lower value $0.119 \pm 0.006$.  However, resummed
QCD~\cite{r11}, in which one uses both $O(\alpha_s^2)$ and next to
leading logarithm corrections in the theoretical expressions, yields
the higher value and better agreement between different
determinations.  The uncertainty is essentially all theoretical, from
scale ambiguities.
This
value is in good agreement with that obtained from the hadronic $Z$
width.  It is somewhat higher than the values obtained by low energy
experiments such as deep inelastic scattering, $\tau$ hadronic decays,
$\Upsilon$ decays, $F_2^\gamma$, and jet production.  It is not clear
whether the uncertainties stated in Table~\ref{tab2} for low energy
determinations are realistic.  There may be larger theoretical
uncertainties.
The value is ({\ref{alsv}) is a reasonable average of the various
measurements with a conservative uncertainty.

One can now extrapolate the couplings to higher energy to see whether
they meet at a point.  This is shown for the standard model (SM) and
for the minimal supersymmetric extension (MSSM) in Figure~\ref{fig5}.
\begin{figure}
\vspace{15cm}
\caption{Extrapolation of the gauge coupling constants in the standard
model and its minimal supersymmetric extension, assuming $\alpha^{-1}
(M_Z) = 127.9 \pm 0.2$, $\sin^2\hat{\theta}_W (M_Z) = 0.2325 \pm
0.0007$, and $\alpha_s (M_Z) = 0.12 \pm 0.01$, updated from
\protect\cite{r6}. }
\label{fig5}
\end{figure}
For the standard model the couplings do not meet.  This means that the
old-fashioned simple grand unified theories such as minimal $SU_5$ are
excluded.  These have also been ruled out for some time by the
nonobservation of proton decay, but the additional evidence is
welcome.  However, the couplings do meet in the supersymmetric
extension.  In Figure~\ref{fig5} it is assumed that all of the new
particles, ({\it i.e.}, the superpartners and extra Higgs particles)
have a common mass $M_{\rm SUSY } = M_Z$.  They still meet within
uncertainties for $M_{\rm SUSY} = 1 \; TeV$ or anywhere between.  The
unification scale $M_X$ is $O(10^{16} \; GeV)$, which is sufficiently
high to suppress proton decay via heavy gauge boson exchange.
However, as we will see there are new sources of proton decay
that may be dangerous.

One can also predict the value of $\sin^2 \hat{\theta}_W (M_Z)$ from
$\alpha^{-1} (M_Z) = 127.9 \pm 0.2$ and $\alpha_s (M_Z) = 0.12 \pm
0.01$.  Recall that this should be $3/8$ if there were no symmetry
breaking.  The predictions are compared with the experimental data in
Figure~\ref{fig6}.  Again, it works beautifully for the MSSM for
reasonable values of $M_{\rm SUSY}$, but not for ordinary grand
unified theories.  The predictions are also shown for the standard
model and the MSSM for two values of $M_{\rm SUSY}$ in
Table~\ref{tab2a}.
\begin{figure}
\vspace{10cm}
\caption{Predictions of $\sin^2 \hat{\theta}_W (M_Z)$ in the ordinary
and supersymmetric grand unified theories compared with the
experimental data. Updated from \protect\cite{r6}.}
\label{fig6}
\end{figure}
\begin{table}\centering
\begin{tabular}{|lll|} \hline
Model & $\sin^2 \hat{\theta}_W (M_Z)$    & $M_X (GeV)$ \\ \hline
Standard Model & $0.2100 \pm 0.0026$ & $ 4.6^{+2.7}_{-1.8} \x 10^{14}$
\\
MSSM $(M_{\rm SUSY} = M_Z)$ & $0.2334 \pm 0.0026$ & $2.5^{+1.3}_{-0.9}
\x 10^{16}$\\
MSSM $(M_{\rm SUSY} = 1\; TeV)$ & $0.2315 \pm 0.0026$ &
$2.0^{+1.0}_{-0.7} \x 10^{16}$ \\
Experiment & $0.2325 \pm 0.0007$ &  -- -- -- \\ \hline
\end{tabular}
\caption{Predictions for $\sin^2 \hat{\theta}_W (M_Z)$ and $M_X$ in
the standard model and the MSSM compared with the experimental value. }
\label{tab2a}
\end{table}

We see that the $\sin^2\theta$ predictions are in excellent agreement
with grand unification of the supersymmetric standard model but not
with non-SUSY unification.  This has actually been known since
1980~\cite{p6,r4,r5}, as shown in Figure~\ref{fig7}.  However, the new
high precision determination of the low energy couplings from LEP make
the agreement especially striking.
\begin{figure}
\vspace{8cm}
\caption{Predictions for $\sin^2\theta_W$ compared with with the
experimental data (solid circles),
for the standard model (open circles) and for the supersymmetric
extension with $M_{\rm SUSY} = M_Z$ (boxes) and $1 \; TeV$ (triangles).}
\label{fig7}
\end{figure}

A number of comments are in order.
\begin{itemize}
\item Is supersymmetry proved?  Absolutely not! The agreement could
very well be an accident, or one can modify ordinary grand unified
theories in some ad hoc fashion.

\item The predictions are independent of the actual GUT group, $SU_5$,
$SO_{10}$, $E_6$, {\it etc}., to an excellent approximation, if the
charge normalization is preserved.

\item At the tree level, $\sin^2 \hat{\theta}_W$ and $M_X$ and the
meeting of the couplings are independent of the number of fermion
families.  (This is not the case for $\alpha_G^{-1} (M_Z)$.)  The reason
is that a fermion family forms a full multiplet of the $SU_5$ group
and thus changes the slope of each $\alpha^{-1}_i$ by the same amount.

\item On the other hand, there is a strong dependence on the number of
Higgs doublets, $N_H$.  That is because a Higgs doublet is part of a
split multiplet associated with heavy partners.  It therefore affects
some couplings differently from other.  In particular it does not
affect the strong coupling.

\item One could improve the $SU_5$ prediction for $\sin^2 \theta_W$ by
increasing the number of Higgs doublets, but that would also decrease
the unification scale, aggravating the proton lifetime problem.

\item Supersymmetry was originally predicted and motivated assuming
that the breaking scale was not too large, {\it e.g.}, $M_Z < M_{\rm
SUSY} \simle 1 \; TeV$.  As we have seen, the coupling constant
predictions are successful for $M_{\rm SUSY}$ anywhere in this range.
One should not, in my opinion, view $M_{\rm SUSY}$ as a parameter to
be fit to the data.  There is no motivation for taking it outside of
this range.

\item The new higher values of $\alpha_s$ favor the smaller values of
$M_{\rm SUSY}$.  One can turn around the logic and use $\alpha +
\sin^2 \hat{\theta}_W$ to predict
\beq \alpha_s (M_Z) \simeq \left\{ \begin{array}{ll} 0.072 \pm 0.001 &
{\rm standard \;\; model} \\
0.125 \pm 0.002 & {\rm MSSM}(M_Z) \\
0.118 \pm 0.002 & {\rm MSSM} (1\; TeV) \end{array} \right. \eeq
Comparing with the results of Table~\ref{tab2a} one sees again
the strong motivation for the MSSM,  especially with the smaller values of
$M_{\rm SUSY}$.

\item There are several additional uncertainties and
corrections~\cite{r9,r12,polonsky}, including low scale uncertainties
associated with the splitting of the masses of the new particles and with
$m_t$, high scale uncertainties from the splitting of the heavy
particles, and from possible non-renormalizable operators~\cite{r13}.
There are also models with intermediate scales, which break in more
than one step to the standard model~\cite{r14}, and epicycle
models~\cite{r15}, involving ad hoc representations split into
superheavy and light pieces, which can significantly affect the
predictions.
\end{itemize}

\section{Complications}

There are several complications which introduce theoretical
uncertainties and limit the precision of the predictions.  At the low
scale we have made the simplifying assumptions that all standard model
particles, including the $t$ and the Higgs scalar, have $m \leq M_Z$,
and that the second Higgs doublet and all of the new SUSY partners
have a common mass scale $M_{\rm SUSY}$.  This is clearly unrealistic.
Ross and Roberts~\cite{r16} have studied threshold effects associated
with a realistic SUSY spectrum including splittings.  They found that
the splittings can have a significant effect, mainly because the
colored superpartners tend to be heavier than the uncolored ones.  A
simple parameterization of this effect is given in \cite{polonsky},
where it is shown that an effective $M_{\rm SUSY}$ can always be
defined, but it may be very different from the actual superpartner
masses.

The high scale is also dangerous~\cite{r12,polonsky,r13}.  We have
implicitly assumed so far that all of the superheavy fermions,
scalars, and vectors have a common mass $M_X$.  In fact, they are
likely to have splittings, leading to high-scale threshold
corrections~\cite{r9}.  Including these\footnote{There are also small
mass-independent discontinuities which depend on the renormalization
scheme~\protect\cite{r9,polonsky}.},
\beqa \alpha_i^{-1}(\mu) &=& \alpha_i^{-1} (M_X) - \frac{b_i}{2\pi}
\ln \left( \frac{\mu}{M_X} \right) + 2 \;{\rm loop} \nonumber \\
&& + \frac{1}{3\pi} \Tr \left[ Q_{iF}^2 \ln \frac{M_X}{M_F} \right] +
\frac{1}{24\pi} \Tr \left[ Q_{iS}^2 \ln \frac{M_X}{M_S} \right] \eeqa
where $Q_{iF,S}$ are respectively the charges of the heavy fermions
and scalars, and $M_{F,S}$ are their mass matrices.  If, for example,
one allows the particles to vary by two orders of magnitude around the
unification scale, $M_X/M_{F,S} \sim 10^{\pm 2}$, the prediction for
$\alpha_s$ may vary by $\Delta \alpha_s (M_Z) = \pm 0.01$.  Thus,
\beq \alpha_s (M_Z) |_{\rm MSSM} \sim 0.12 \pm 0.01.\eeq
All values in this range are compatible with the experimental value,
but due to the theory uncertainties a more precise measurement is
useless in this context.  Similarly, $M_{\rm SUSY}$ cannot be
determined in this way.  On the other hand, in the nonsupersymmetric
case one predicts
\beq \alpha_s (M_Z) |_{\rm SM} \sim 0.072, \eeq
which is incompatible with the data for any reasonable uncertainties.
There are other high-scale corrections associated with possible
non-renormalizable operators~\cite{r13} such as
\beq \L \sim \frac{1}{M_P} \Tr (F_{\mu \nu} F^{\mu \nu} \phi).\eeq
All of these corrections are studied in detail in
reference~\cite{polonsky}.

The above high and low scale corrections are to be viewed as
uncertainties in the simple two-scale grand unified theories in which
there is indeed a desert between the low and high scales.  One can
consider more drastic modifications, such as intermediate-scale
models~\cite{r14}, in which the grand unified theory breaks in two or
more steps to the standard model, or models involving ad hoc new split
multiplets of fermions and scalars~\cite{r15}.  These have sufficient
freedom that one can bring the nonsupersymmetric case into agreement
with unification.  Such models are logical possibilities but have
little predictive power.

\section{Implications of SUSY Unification}

\subsection{Proton Decay}

In the ordinary nonsupersymmetric grand unified theories, such as
those based on $SU_5$, $SO_{10}$, and $E_6$ one predicts a unification
scale
\beq M_X \sim (2 - 7) \x 10^{14} \; GeV \eeq
from the observed values of $\alpha$ and $\alpha_s$, as well as the
         incorrect prediction for $\sin^2\theta_W$.   From the
diagrams in Figure~\ref{fig2} one can predict the proton lifetime into
    $e^+\pi^0$.       There are large theoretical uncertainties,
from the value of $M_X$ and also from the hadronic matrix element.
One finds~\cite{p6}
\beqa \tau_{p \ra e^+e^0} &\sim & \frac{M_X^4}{\alpha_G^2 m_p^5}
\nonumber \\
&& \sim  10^{31 \pm 0.7} \left( \frac{M_X}{4.6 \x 10^{14}}
\right)^4 yr \nonumber \\
&& \sim  10^{31 \pm 1} yr,\eeqa
which is low compared to the experimental limit~\cite{barloutaud}
\beq \tau_{p \ra e^+\pi^0} > 10^{33} yr.\eeq
Thus, these models seem to be excluded on the basis of proton decay.
It is interesting that a few years ago the theoretical prediction
$\tau_{p \ra e^+\pi^0} \sim 10^{29 \pm 3}$ yr was shorter but with a
larger uncertainty.  The change is due to the exponential dependence
of the lifetime prediction on $\alpha_s (M_Z)$.  The new higher values
of $\alpha_s (M_Z)$ predict a longer lifetime.  Therefore, while the
new precision measurements tighten the disagreement with the $\sin^2
\theta_W$ predictions, they have weakened the discrepancy based on
proton decay.

\subsection{SUSY-GUT}

In the supersymmetric grand unified theories, on the other hand, one
has a much higher prediction
\beq M_X \sim 2.5 \x 10^{16} \; GeV \eeq
for the unification scale.  This strongly suppresses the decay rate
via the gauge boson exchange to a comfortably safe \beq \tau_{p \ra
e^+ \pi^0} \sim 3 \x 10^{38 \pm 1} \; yr.\eeq However, SUSY-GUTs have new
dimension $d=5$ operators~\cite{r17} mediated by the exchange of a
heavy Higgsino, as in Figure~\ref{fig8}.  The basic baryon
number-violating process
\begin{figure}
\vspace{5cm}
\caption{Diagram for a proton decay in supersymmetric grand unified
theories.  The basic operator is due to the heavy Higgsino exchange on
the right-hand side of the diagram.}
\label{fig8}
\end{figure}
\beq \tilde{q} \tilde{q} \ra \bar{q} \bar{l} \eeq
must be dressed with the exchange of a light gaugino, $\tilde{w}$,
$\tilde{z}$, $\tilde{g}$, in order to generate a diagram for $qq \ra
\bar{q} \bar{l}$.  Since the process is mediated by the exchange of a
heavy fermion the lifetime goes like $\tau_p \sim M_X^2$ rather than
$M_X^4$, so it is extremely dangerous.  It tends to produce decays
which change generations, such as into $\bar{\nu} K^+$.  One
predicts~\cite{r17}
\beq \tau_{p \ra \bar{\nu} K^+} \sim 10^{29 \pm 4},\eeq
which is only marginally compatible with the current experimental
limit~\cite{barloutaud}
\beq \tau_{p \ra \bar{\nu} K^+} > 10^{32} yr.\eeq

Nath and Arnowitt~\cite{r17} have done a detailed study of the $d=5$
constraints in supersymmetric grand unified theories.  They find that
the so-called ``no-scale'' models are excluded by proton decay.  More
general supersymmetric models are still viable, but only if they
satisfy constraints on the low energy spectrum, such as $m_t < 175$
GeV; $m_{\tilde{g}} < m_{\tilde{q}}$; $M_H < M_Z$; and a chargino and
two neutralinos $< 100 \; GeV$.  It should be commented that the $d =
5$ operators may be absent in certain theories which are not true
grand unified theories, such as some string theories and flipped $SU_5
\x U_1$.

\section{Neutrino Mass}

Grand unified theories often have interesting implications for
neutrino mass.  Typically, they lead to a seesaw
prediction~\cite{r18}, in which the light neutrino masses are of order
\beq m_{\nu_i} \sim c_i \frac{m_{u_i}^2}{M_{N_i}},\eeq
where $u_i = u$, $c$, $t$ are the light charged $2/3$ quarks and $c_i
\sim 0.05 - 0.4$ are coefficients which depend on the radiative
corrections or the running of the masses from the high scale down to
the low scale.  The $N_i$ are heavy majorana neutrinos.  Furthermore,
simple grand unified theories often predict the relation
\beq V_{\rm lept} \simeq V_{\rm CKM} \label{eq:b} \eeq
between the lepton and quark mixing matrices.

In the ``old'' supersymmetric grand unified theories which were
prevalent before the days of superstring theories one often introduced
very large Higgs representations, such as a $126$-plet of $SO_{10}$.
In such models these Higgs can generate large majorana masses,
typically
\beq M_{N_i} \sim (10^{-2} - 1) M_X, \eeq
which implies the very small neutrino masses
\beqa m_{\nu_e} & \simle & 10^{-11} \; eV \nonumber \\
      m_{\nu_\mu}& \sim   &(10^{-8} - 10^{-6}) \; eV \nonumber \\
      m_{\nu_\tau}& \sim   &(10^{-4} - 1 ) \; eV.\eeqa
These masses are sufficiently small that if one wishes to satisfy the
MSW solution to the solar neutrino model one must postulate~\cite{r14}
$\nu_e \ra \nu_\tau$.  If the equation (\ref{eq:b}) is satisfied then
one expects that this will be in a disfavored small-angle region, as
shown in Figure~\ref{fig9}.
\begin{figure}
\vspace{10cm}
\caption[]{Regions for the neutrino mass and mixing parameters favored
by solar neutrino data and other experiments, compared with the
typical predictions of various grand unified theories.
The two small regions marked ``combined'' are allowed by the
combination of the Kamiokande, Homestake, and GALLEX results~\cite{r14}.}
\label{fig9}
\end{figure}

On the other hand, in the string-inspired modern versions of
supersymmetric theories it is difficult to introduce large Higgs
representations.  One therefore expects
\beq M_{N_i} = 0 \eeq
in the simplest versions at the lowest order.  However, it is quite
possible to have~\cite{r20}
\beq m_{N_i} \sim 10^{-4} M_X \sim 10^{12} \; GeV \eeq
generated by non-renormalizable gravity-induced operators, which may
well survive the compactification.  In this case one would typically
expect
\beqa m_{\nu_e} & \simle & 10^{-7} \; eV \nonumber \\
m_{\nu_\mu} &\sim& 10^{-3} \; eV \nonumber \\
m_{\nu_\tau} &\sim & (3 - 21) \; eV.\eeqa
One could easily have $\nu_e \ra \nu_\mu$ for the solar neutrinos and
observable $\nu_\mu \ra \nu_\tau$ oscillations in the laboratory.
Furthermore, $m_{\nu_\tau}$ may be in the cosmologically-relevant
range of a few eV.  Such models have considerable flexibility in the
masses and do not make any firm predictions about the mixing angles.

Finally, intermediate scale models (ordinary grand unified theories
breaking in two stages to the standard model) give results similar to
the ``new'' SUSY models if the intermediate scale is of order
$10^{12}\;GeV$.  One actually expects a somewhat lower value, $10^{10}
\;GeV$, in the simplest $SO_{10}$ models compatible with the coupling
constants.

Let me briefly comment on a few other implications of supersymmetric
theories.
\begin{itemize}
\item The Supersymmetric Spectrum.  Ross and Roberts~\cite{r16} have
considered realistic SUSY spectra based not only on unification but
criteria of naturalness and $m_b$.  Nath and Arnowitt~\cite{r14} have
considered the constraints from proton decay.  In both cases they
predict rather low scales for the superpartners, such as
\beqa m_{\tilde{\gamma}, \tilde{w}, \tilde{z}} &\sim & 100 \; GeV
\nonumber \\
m_{\tilde{g}, \tilde{q}} & \sim & (300 - 500) \; GeV.\eeqa

\item Many simple grand unified theories predict
\beq m_b = m_\tau \eeq
at the GUT scale. Including running effects mainly due to gluon
exchange   this is renormalized to the prediction
\beq m_b \sim (4 - 5) \; GeV \eeq
at the low scale.  However, such models also make the bad prediction
\beq \frac{m_d}{m_s} = \frac{m_e}{m_\mu}, \eeq
which fails by an order of magnitude.

\item The connection of these ideas to superstring theories is still
rather strained~\cite{r20,r21}.  In the simplest superstring
compactifications one would expect a gauge group $G \subset E_6$ to
emerge at the compactification scale $M_C \sim 10^{18} \; GeV$.  In
general one would not expect this to be a simple GUT group.  For
example, it could be just the standard model itself.  Even for schemes
for which $G = SU_5$, $SO_{10}$, $E_6$ one typically expects $M_X \sim
M_C$, while the data are suggesting that $M_X$ is one or two orders of
magnitude lower.  It is possible~\cite{r21} that the compactification
does result in the standard model group, but that there are large
threshold corrections from massive modes which cause the couplings to
cross at an effective scale lower than $M_C$.  However, there are no
compelling or realistic models of this, and other GUT effects, such as
the $m_b$ and proton decay predictions, may be lost.

\item There are potentially interesting implications of unification
for the large scale structure of the universe, but so far this is very
speculative.

\item What about baryogenesis?  The successful prediction of the
baryon asymmetry of the universe was a great success of the ordinary
GUTs~\cite{p4}.  Now, however, there is the major complication from
baryon number violation at the electroweak scale~\cite{p3}.  This may
wash out any baryon asymmetry created earlier unless it is either
enormous or leads to a non-zero value of $B-L$. \end{itemize}

\section{Conclusions}

\begin{itemize}
\item We expect $B$ violation to occur in nature at some level.
\item Proton decay has not yet been observed.
\item In the future, one can increase the sensitivity of proton decay
searches by one or two orders of magnitude.
\item There are strong motivations to continue the search for $B$
violation from the coupling constant unification in the MSSM.  This
could well be an accident, but it may also be a hint that the simple
ideas of the grand desert and supersymmetry are on the right track.
Such schemes would have many interesting implications, such as for
proton decay $p \ra \bar{\nu} K^+$, neutrino mass, $m_b$, the SUSY
spectrum, cosmology, {\it etc}.
\end{itemize}

\newcommand{\tn}{n}
\newcommand{\chc}{\v{c}}
\newcommand{\du}{u}
\newcommand{\aca}{a}

\end{document}